\documentclass[review]{elsarticle}

\usepackage{hyperref}
\usepackage{amsmath,amssymb,amsfonts}
\usepackage{algorithmic}
\usepackage{ntheorem}
\usepackage{caption}
\usepackage{graphicx}
\usepackage{textcomp}
\usepackage{indentfirst}
\usepackage{float}
\usepackage{bm}
\usepackage{tagging}
\usepackage{amsfonts,amssymb}
\usepackage{amsmath}
\usepackage{tablists}
\usepackage{subfigure}
\usepackage{stfloats}
\usepackage{color}
\usepackage{appendix}
\DeclareMathAlphabet\mathbfcal{OMS}{cmsy}{b}{n}

\journal{Journal of \LaTeX\ Templates}

\begin{document}
\captionsetup[figure]{labelfont={bf},labelformat={default},labelsep=period,name={Fig.}}

\begin{frontmatter}

\title{Weighted Mean and Median graph Filters with Attenuation Factor for Sensor Network  \tnoteref{mytitlenote} }
\tnotetext[mytitlenote]{This work has been supported by National Natural Science Foundations of China (No.62071242)}

\author{Zirui~Ge, Haiyan~Guo, Tingting~Wang, Zhen~Yang\corref{mycorrespondingauthor} }
\address{School of Communication and Information Engineering, Nanjing University of Posts and Telecommunications, Nanjing 2100023, China}
\fntext[myfootnote]{E-mail address: yangz@njupt.edu.cn (Z. Yang)}

\cortext[mycorrespondingauthor]{Corresponding author}


\begin{abstract}
This paper proposes a weighted attenuation k-hop graph, which depicts the spatial neighbor nodes with their hops from the central node. Based on this k-kop graph, we further propose a node selecting graph, which selects temporal neighbor nodes of multiple instances of the central node. With this node selecting graph, we propose a graph mean filter. In addition, we also apply the proposed node selecting graph to the median filter. Finally, the experimental results show that the proposed mean filter performs better than the original median filter in the signal denoising polluted by white noise and the median filter using node selecting graph also has better performance than the original median filter.
\end{abstract}

\end{frontmatter}

\section{Introduction}
\par Sensor networks often operate in harsh environments, and the measured values of physical variables may be seriously damaged \cite{Ref1,Ref2,Ref3}. This requires efficient and straightforward tools to reduce the noise of measurement data, which is the research motivation of this paper.

\par Due to the excellent performance of graph signal processing on irregular data sets, such as sensor networks and social networks, graph signal processing technology has attracted much attention from researchers \cite{Ref4,Ref5,Ref6}. Many classical signal processing theories have been extended to graph signal processing, including graph Fourier transform, graph filter, graph filter bank, and so on. Among them, graph filter plays a critical role which is applied to the noise reduction of graph signals and applied to the fields of graph signal compression, classification, interpolation, and clustering. GSP technology is also used for sensor network signal processing \cite{Ref7,Ref8,Ref9}.
\par Graph shift operator is the basic module of graph signal processing. In \cite{Ref10}, the graph shift operator is the adjacency matrix of the graph and simple justification of such a choice is presented. It is defined in \cite{Ref11} as the translation on graph via generalized revolution with a delta centered at vertex $n$. On the other hand, the polynomial combination of graph shift operators constitutes the representation of graph filter, such as graph IIR filter and graph FIR filter \cite{Ref12,Ref13,Ref14}. 

\par In addition to the above linear graph filter in the form of shift operator polynomial, \cite{Ref15,Ref16} proposed the application of nonlinear filter. The simplest nonlinear filter is the graph median filter, first proposed by \cite{Ref17,Ref18}. However, this filter is mainly applied to the static graph model, and the current graph signal is constructed under the time-varying graph model \cite{Ref19}. In \cite{Ref20, Ref21}, the author proposed the application of k-hop graphs and recursive median filters to time-varying graph signals. However, the k-hop graph considers the neighbors of different node hops have the same importance. In practice, the farther the path from the central node is, the less influence the neighbor has on the central node. In addition, the main disadvantage of median filter is that each vertex in the graph signal is replaced by the median in its neighborhood, which is likely to be a noise signal under high noise density \cite{Ref22}. 

\par Aiming at the shortcomings of the graph median filter, we propose a graph mean filter based on a node selecting graph. We first propose a weighted attenuation k-hop graph, which provides an attenuation factor to choose the neighbor nodes around the central vertex in the space dimension. In the time dimension, we add another attenuation parameter to choose neighbor nodes of multiple instances of the central node. Our node selecting graph is constructed by this k-hop graph and the temporal attenuation parameter. Then, we propose the graph mean filter based on this node selecting graph and apply this node selecting graph to graph median filter. To our best knowledge, this is the first time to apply the mean filter to the time-varying graph signal. Different from the mean filter in image processing, it selects the required pixels according to the spatial structure of pixels. The mean graph filters depend on the local characteristics of the graph shift operator.

\par The main contributions of this paper are summarized as follows:

\par 1)	We propose a weighted node selecting graph and use it to improve the recursive median graph filter. The weighted node selecting graph considers the nodes at multiple moments before and after the central node, instead of the non-weight attenuation graph in \cite{Ref21}, which only considers the two times before and after the central node.

\par 2)	We further propose a mean filter for time-varying graph signal processing using our node selecting graph. Compared with the median filter, it has less computational complexity and higher performance in the case of white noise.

\par The outline of the paper is as follows. Section II introduces the basic concepts of graph signal processing and the k-hop graph. Section III shows how to obtain the weighted node selecting graph. In Section IV, we discuss the graph mean filter and median filter with weighted node selecting graph. Our simulation results are provided in Section V, while Section VI concludes the paper.

\section{Preliminaries}
\subsection{Graph Signals.} Consider an graph $G_{\mathcal{G}}=\left( V,E \right) $ with $N$ nodes and $M$ edges, where $V=\left[ \begin{matrix}
	v_1&		v_2&		\cdots&		v_N\\
\end{matrix} \right] $ represents $N$ vertices set and $E$ $M$ edges set. The underlying structure of $G_{\mathcal{G}}$ is described by the adjacency matrix $\boldsymbol{A}_{\mathcal{G}}\in R^{N\times N}$ or by Laplacian matrix $\boldsymbol{L}_{\mathcal{G}}=\boldsymbol{D}_{\mathcal{G}}-\boldsymbol{A}_{\mathcal{G}}$, where $\boldsymbol{D}_{\mathcal{G}}$ is the diagonal degree matrix. For an undirected graph $G_{\mathcal{G}}$, every edge between $v_i$ and $v_j$ is same as the edge in $v_j$ and $v_i$, and the Laplacian matrix is symmetric.
\subsection{Time-vertex graphs.}
\par The Kronecker product is used to describe the joint time-vertex domain. The Kronecker product graph is defined as $G_J=G_{\mathcal{T}}\otimes G_{\mathcal{G}}$, where graph $G_{\mathcal{T}}$ denotes a directed line graph with $T$ time instants. A time-varying signal residing on a product graph $G_J$ can be represented by $\mathbf{X}=[x_1,...,x_T]_{N×T}$, where $x_t$ is a static graph signal on graph $G_j$ at instant $t$, $1\leqslant t\leqslant T$. Now, $x_{t,i}$ is the $i$th node value of the static graph signal at the instant $t$, i.e., the $(i,t)$th entry of the signal matrix $\boldsymbol{X}$. The Laplacian matrix of the product graph $G_J$ is defined as $L_J=L_{\mathcal{T}}\otimes L_{\mathcal{G}}$, where $\boldsymbol{L}_{\mathcal{T}}$ refers to the Laplacian matrix of the directed line graph $G_j$ \cite{Ref19}.
 \subsection{The k-hop Graph.}

\par Let $\boldsymbol{A}_{\mathcal{S}}$ denote the adjacency matrix of the undirected line graph that models the time correlation of the signal values. All elements of $\boldsymbol{A}_{\mathcal{S}}$ are zero, except those of the two off-diagonals, which are ones. Let $\boldsymbol{A}_{\mathcal{L}}$ denote the logical adjacency matrix of the a graph 
$G_{\mathcal{G}}$, and ${\boldsymbol{A}_{\mathcal{L}}}_{\left[ i,j \right]}=1$ for $\boldsymbol{L}_{\mathcal{G}}\ne 0$, ${\boldsymbol{A}_{\mathcal{L}}}_{\left[ i,j \right]}=0$ for $\boldsymbol{L}_{\mathcal{G}}=0$. Let define another adjacency matrix $\boldsymbol{A}_{G,K}$ representing another graph, with the same vertices as $G_{\mathcal{G}}$, called the K-hop graph. For unweighted graphs, $\boldsymbol{A}_{G,K}$ can be obtained as follows:
$$
\boldsymbol{A}_{G,K}=\left( \left( \sum_{k=1}^K{\boldsymbol{A}_{G}^{k}} \right) >\boldsymbol{0}_N \right) -\boldsymbol{I}_N,
$$
where $\boldsymbol{I}_N$ denotes the identity matrix of size $N$, and $\boldsymbol{0}_N$ is the all zero size $N $ matrix. The operator $>$ is an element-wise logical operator between the values of the two matrices. 
\par The strong product graph’s adjacency matrix is defined as:
$$
\begin{aligned}
	&\boldsymbol{A}_{SP}=\boldsymbol{I}_T\otimes \boldsymbol{A}_{G,K}+\boldsymbol{A}_T\otimes \left( \boldsymbol{A}_{G,K}+\boldsymbol{I}_N \right)\\
	&=\left[ \begin{matrix}
	\boldsymbol{A}_{G,K}&		\left( \boldsymbol{I}_N+\boldsymbol{A}_{G,K} \right)&		...&		0&		0\\
	\left( \boldsymbol{I}_N+\boldsymbol{A}_{G,K} \right)&		\boldsymbol{A}_{G,K}&		...&		0&		0\\
	0&		\left( \boldsymbol{I}_N+\boldsymbol{A}_{G,K} \right)&		...&		0&		0\\
	\vdots&		\vdots&		\vdots&		\vdots&		\vdots\\
	0&		0&		0&		\left( \boldsymbol{I}_N+\boldsymbol{A}_{G,K} \right)&		\boldsymbol{A}_{G,K}\\
\end{matrix} \right]\\
\end{aligned},
$$
where
$$
\boldsymbol{A}_{\mathcal{S}}=\left[ \begin{matrix}
	0&		1&		0&		\\
	1&		0&		\ddots&		0\\
	0&		\ddots&		\ddots&		1\\
	&		0&		1&		0\\
\end{matrix} \right] .
$$
The strong product graph depicts the neighbor nodes of a central node at different time in space and time dimension.
\section{The Weighted Node Selecting Graph}
In this section, we discuss how to improve the k-hop graph and the node selecting graph. In \cite{Ref21}, the node selecting graph selects the same nodes for a central node $x_{t,i}$, at instance $t-1$, $t$ and $t-1 $. Intuitively, as the distance from instance $t$ increases, the relationship of nodes in different instances for the node in instance $t$ weakens. We believe that the neighbor nodes of $x_{t,i}$ in instance $t$ are different from the nodes in instances  $t-1$ and $t+1$. Besides, the node selecting graph in \cite{Ref21} only consider the neighbor nodes in instances $t-1$ and $t-1$, and more instances should be included.
\par We first add a spatial attenuation parameter $\beta $ and threshold $\boldsymbol{\gamma }$ to improve the k-hop graph $\boldsymbol{A}_{G,K}$. When the relationship value of the neighbor node is lower than the threshold $\boldsymbol{\gamma }$, we believe that the neighbor nodes have no direct relationship with the central node, and set the value to 0, otherwise it is 1. The improved $\boldsymbol{A}_{G,K}^{p}$ is represented as 
$$
\boldsymbol{A}_{G,K}^{p}=\left( \left( \sum_{k=1}^K{\beta ^k\boldsymbol{A}_{G}^{k}} \right) >\boldsymbol{\gamma }_N \right) -\boldsymbol{I}_N,
$$
where $K$ and $\boldsymbol{\gamma }$ the specify the size of the vertex spatial neighbourhood. 
The selected node set $\mathcal{N} _s$ of a center node $v_i$ with $\boldsymbol{A}_{G,K}$ and $\boldsymbol{A}_{G,K}^{p}$ are shown in Fig.1.

\begin{figure}[H]
\centering
\subfigure[]{
\includegraphics[width=1in]{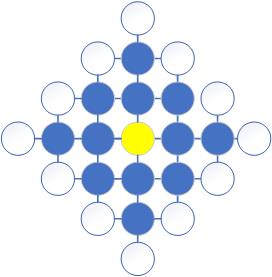}
}
\subfigure[]{
\includegraphics[width=1in]{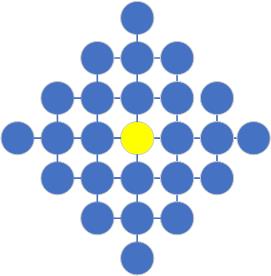}
}
\caption{The selected node sets $\mathcal{N} _s$ of a center node. The yellow nodes are the center nodes, and blue nodes are the selected neighbour nodes. (a) the selected nodes with $\boldsymbol{A}_{G,K}^{p}$. (b) the selected nodes with $\boldsymbol{A}_{G,K}$.}
\end{figure}
\par Furthermore, we consider more moments before and after the central time. Meanwhile, to control the quantities of neighbor nodes, we add a time attenuation parameter $\alpha$. Similarly, when the relationship value of the neighbor node is lower than the threshold $\mathbf{\gamma }$, the neighbor value is set to 0, otherwise 1. The improved $\boldsymbol{A}_{SP}^{p}$ is represented as 
$$
\begin{aligned}
&\boldsymbol{A}_{S P}^{p}=\boldsymbol{I}_{T} \otimes\left(\boldsymbol{A}_{G, K}^{p}+\boldsymbol{I}_{N}\right)+\boldsymbol{A}_{T}^{p} \otimes\left(\boldsymbol{A}_{G, K}^{p}+\boldsymbol{I}_{N}\right) \\
&{\left[\begin{array}{cccc}
\boldsymbol{I}_{N}+\boldsymbol{A}_{G, K}^{p} & \alpha\left(\boldsymbol{I}_{N}+\boldsymbol{A}_{G, K}^{p}\right)>\boldsymbol{\gamma}_{N} & \cdots & \boldsymbol{0} \\
\alpha\left(\boldsymbol{I}_{N}+\boldsymbol{A}_{G, K}^{p}\right)>\boldsymbol{\gamma}_{N} & \boldsymbol{I}_{N}+\boldsymbol{A}_{G, K}^{p} & \cdots & \boldsymbol{0} \\
\vdots & \alpha\left(\boldsymbol{I}_{N}+\boldsymbol{A}_{G, K}^{p}\right)>\boldsymbol{\gamma}_{N} & & \boldsymbol{0} \\
\alpha^{M}\left(\boldsymbol{I}_{N}+\boldsymbol{A}_{G, K}^{p}\right)>\boldsymbol{\gamma}_{N} & \vdots & \cdots & \alpha^{M}\left(\boldsymbol{I}_{N}+\boldsymbol{A}_{G, K}^{p}\right)>\boldsymbol{\gamma}_{N} \\
\boldsymbol{0} & \alpha^{M}\left(\boldsymbol{I}_{N}+\boldsymbol{A}_{G, K}^{p}\right)>\boldsymbol{\gamma}_{N} & \cdots & \vdots \\
\vdots & \vdots & \vdots & \alpha\left(\boldsymbol{I}_{N}+\boldsymbol{A}_{G, K}^{p}\right)>\boldsymbol{\gamma}_{N} \\
\boldsymbol{0} & \boldsymbol{0} & \boldsymbol{0} & \boldsymbol{I}_{N}+\mathbf{A}_{G, K}^{p}
\end{array}\right]_{N T \times N T}}
\end{aligned},
$$
where 
$$
\begin{aligned}
&\boldsymbol{A}_{T}^{p}=\left[\begin{array}{ccccccc}
0 & 1 & \cdots & 1 & & & \\
1 & 0 & 1 & \ddots & \ddots & & \\
\vdots & \ddots & 0 & \ddots & & 1 & \\
1 & \ddots & \ddots & \ddots & 1 & \ddots & 1 \\
& \ddots & \ddots & 1 & 0 & 1 & \vdots \\
& & 1 & \ddots & 1 & 0 & 1 \\
& & & 1 & \cdots & 1 & 0
\end{array}\right] \\
&=\text { Toeplitz }([0, \text { ones }(1, M), \text { zeros }(1, N-M-1)])
\end{aligned},
$$
and $M$ is time parameter that controls the number of neighbor instances of the central time.
\par Given a node $x_{t,i}$ in the time-vertex graph signal with $T$ instances, its selected neighbor nodes set $\mathcal{N} $ are shown in Fig.2. Observe from the Fig.2, we can obtain that our nodes selecting graph chooses different number of nodes at different instances, and as the instance from the central node increases, fewer nodes are selected.
\begin{figure}[H]
\centering
\subfigure[]{
\includegraphics[width=3.5in]{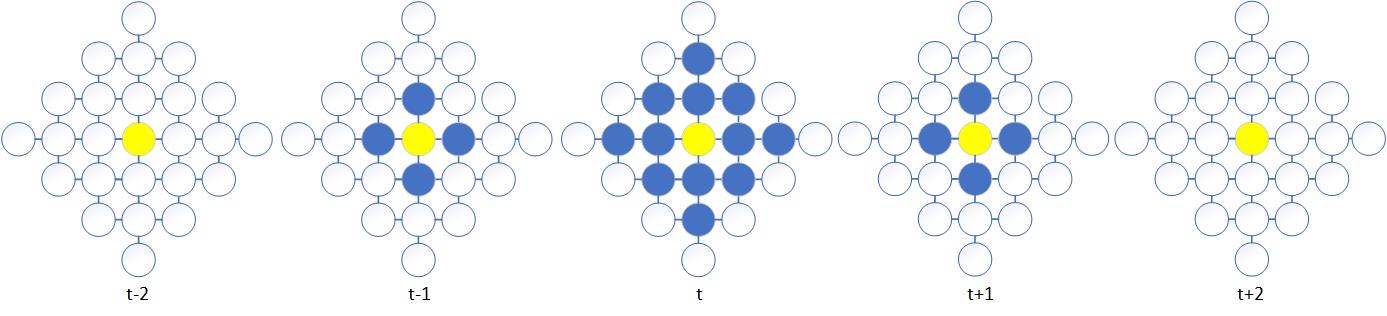}
}
\subfigure[]{
\includegraphics[width=3in]{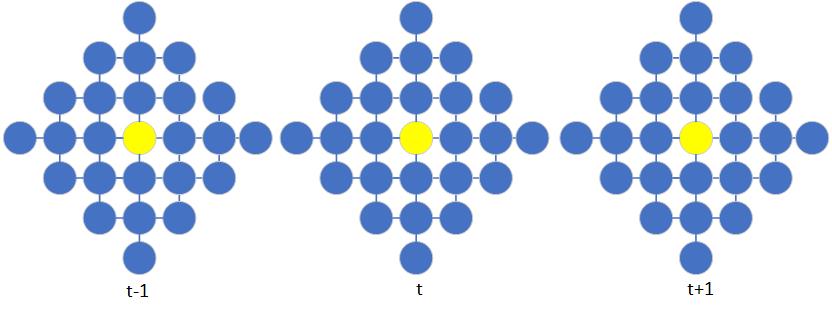}
}
\caption{The selected node sets $\mathcal{N} _s$ of a center node. The yellow nodes are the center nodes, and blue nodes are the selected neighbour nodes. (a) the selected nodes with $\boldsymbol{A}_{SP}^{p}$. (b) the selected nodes with $\boldsymbol{A}_{SP}$.}
\end{figure}
\section{ Graph Mean Filters and Improved Graph Median Filters with Weighted Nodes selecting Graph}
The mean filter operates on a set of signal values for a given time-vertex node $x_{t,i}$. The node set is the neighbourhood of values defined by the adjacency matrix $\boldsymbol{A}_{SP}$. To better depict the node set of the central node $x_{t,i}$, we denote the set by $\mathcal{N} (i,t;K,M)$, where $K$ and $M$ are the spatial and time hyperparameter. For convenience in further development, we partition $\mathcal{N} (i,t;K)$ into disjoint subsets as follows:
$$
\mathcal{N} (i,t;K,M)=\bigcup_{l=1}^M{\left( \mathcal{N} (t-l;i,K)\bigcup{\mathcal{N} (t+l;i,K)} \right)}
$$
where $\mathcal{N} (t-l;i,K)$ and $\mathcal{N} (t+l;i,K)$ contains the time-vertex nodes at instance $t-l$ and $t+l$. We denote the set of signal values defined over these nodes as $g\left( \mathcal{N} \right) $, where $\mathcal{N} $ is the node set, and denote $y\left( i,t \right) $ as output of the mean filter.
\par The filter operates sequentially from time $t=1$ till $t=T$ as follows:
\par 1)	When $t<M$, the output is given by 
$$
y\left( i,t \right) =MEAN\left( g\left( \bigcup_{l=1}^M{\mathcal{N} _S(t+l;i,K)} \right) ,g\left( \bigcup_{l=1}^{t-1}{\left( \mathcal{N} _S(t-l;i,K) \right)} \right) \right) 
$$
\par 2)	When $M\leqslant t\leqslant T-M$, the output is given by 
$$
y\left( i,t \right) =MEAN\left( \mathcal{N} _{S,T}(i,t;K,M) \right) 
$$
\par 3)	When $T-M<t$, the output is given by:
$$
y\left( i,t \right) =MEAN\left( g\left( \bigcup_{l=1}^M{\mathcal{N} _S(t-l;i,K)} \right) ,g\left( \bigcup_{l=1}^{T-t}{\left( \mathcal{N} _S(t+l;i,K) \right)} \right) \right) 
$$
The median filter with weight the node selecting graph performs in the same way as in \cite{Ref21} except that the selected vertices are different. 

\par Note that, the median filter can only be implemented by nodes, and the mean filter can be executed in batches.
\par Given a time-vertex graph signal $
\boldsymbol{X}=\left[ \begin{matrix}
	x_1&		\cdots&		x_T\\
\end{matrix} \right] ^{N\times T}
$,
$
\boldsymbol{x}=vec\left( \boldsymbol{X} \right) 
$
, the graph mean filter output of $\boldsymbol{x}$ is denoted as 
$$
y=\mathrm{diag}\left( c \right) \mathbf{A}_{SP}^{p}x
$$
where $
c=\left[ \begin{matrix}
	1/d_1&		1/d_2&		\cdots&		1/d_{NT}\\
\end{matrix} \right] ,
$
$d_i$ is the diagonal entries of degree matrix $\boldsymbol{D}_{SP}^{p}$ of $\boldsymbol{A}_{SP}^{p}$.
\section{ Experimental results}
\par We will conduct signal denoising experiments on two sensor network data sets. Noise signals will be tested at different input SNR. The input SNR is calculated as follows:	
$$
SNR=10\log \left( \frac{\left\| \boldsymbol{X} \right\| _{F}^{2}}{\left\| \boldsymbol{X}-\boldsymbol{Y} \right\| _{F}^{2}} \right) 
$$
where $\boldsymbol{X} \in \mathbb{R}^{N\times T}$ is the original signal, $\boldsymbol{Y}=\boldsymbol{X}+\boldsymbol{N}$ is noisy signal, $\boldsymbol{N}\in \mathbb{R} ^{N\times T}$ is additive white Gaussian noise.
\par The two real-world datasets used in our experiments are:
\par 1)	Global sea-level pressure dataset. The global sea-level pressure dataset was published by the Joint Institute for the Study of the Atmosphere and Ocean. The selected 500 nodes are on the world from $60^\circ$ south to  $10^\circ$ north and from $110^\circ$ west to $170^\circ$ west. Each node collects the global sea-level pressure signals over a time period of 500. A graph is constructed by the 5-nearest neighbors algorithm, shown in Fig.1(a). 
\par 2)	The Sea Surface Temperature dataset. The sea surface temperature dataset was published by the Earth System Research Laboratory. 100 nodes are selected over a time period of 120. The selected nodes are on the Pacific Ocean from $30^\circ$ south to $60^\circ$ south and from $170^\circ$ west to $90^\circ$ west. A graph is constructed by the 5-nearest neighbors algorithm, shown in Fig.3(b). 
\begin{figure}[H]
\centering
\subfigure[]{
\includegraphics[width=3.5in]{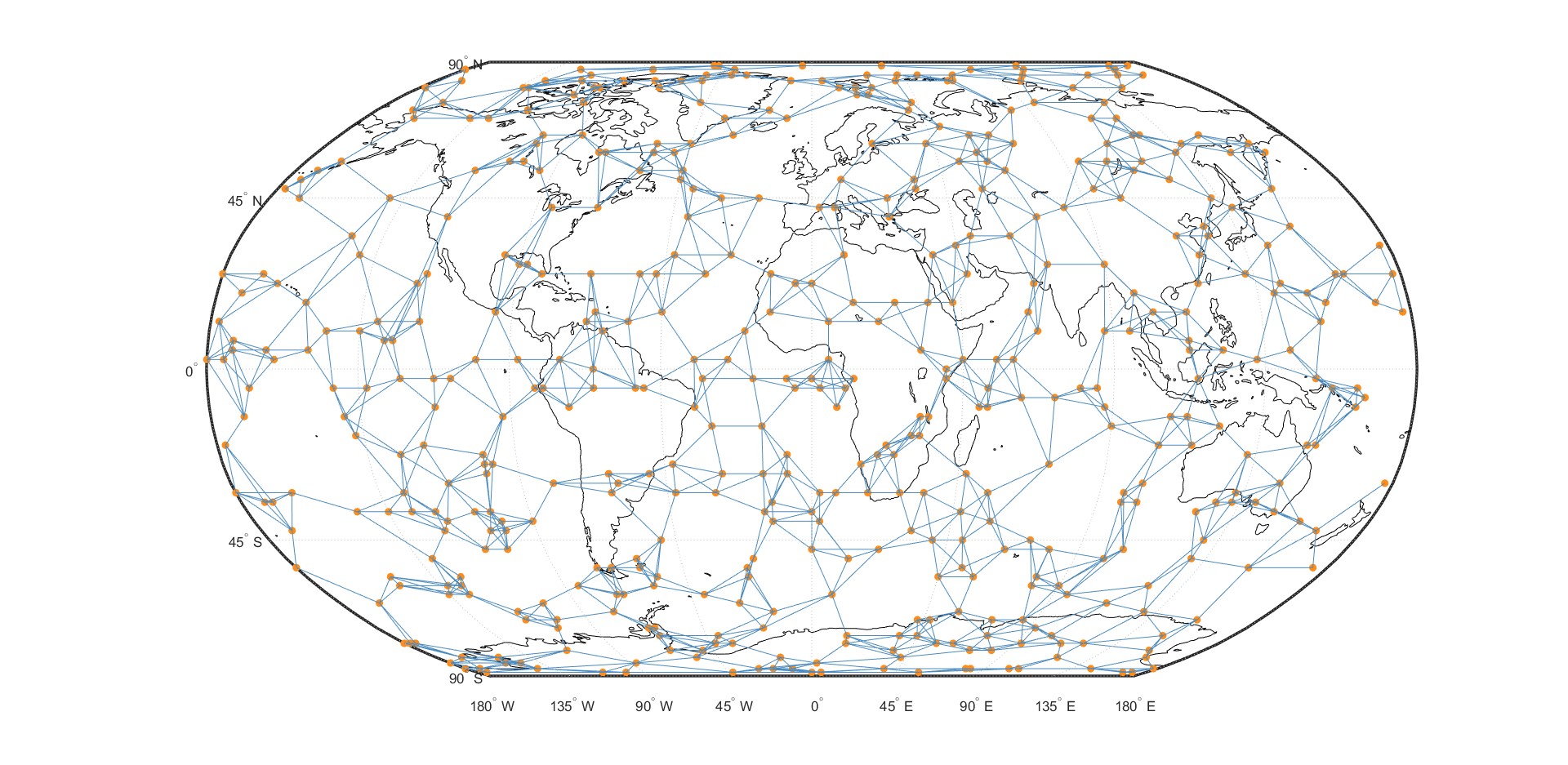}
}
\subfigure[]{
\includegraphics[width=3in]{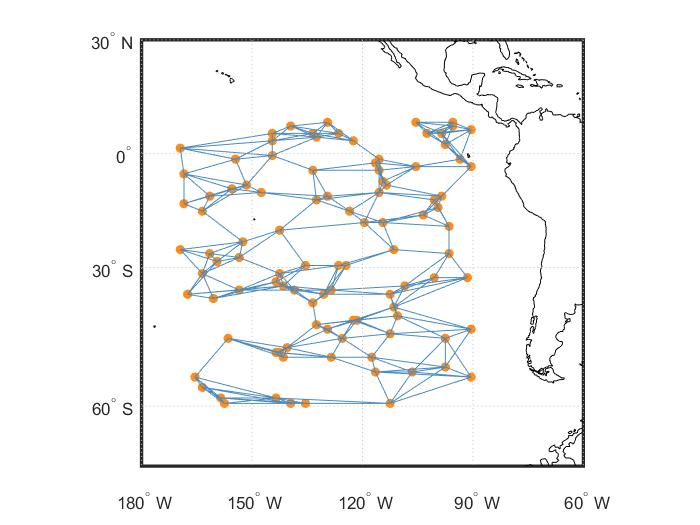}
}
\caption{Sensor networks with two datasets. (a): Global sea-level pressure dataset. (b): The Sea Surface Temperature dataset. }
\end{figure}

\begin{figure}[H]
\centering
\subfigure[]{
\includegraphics[width=3.3in]{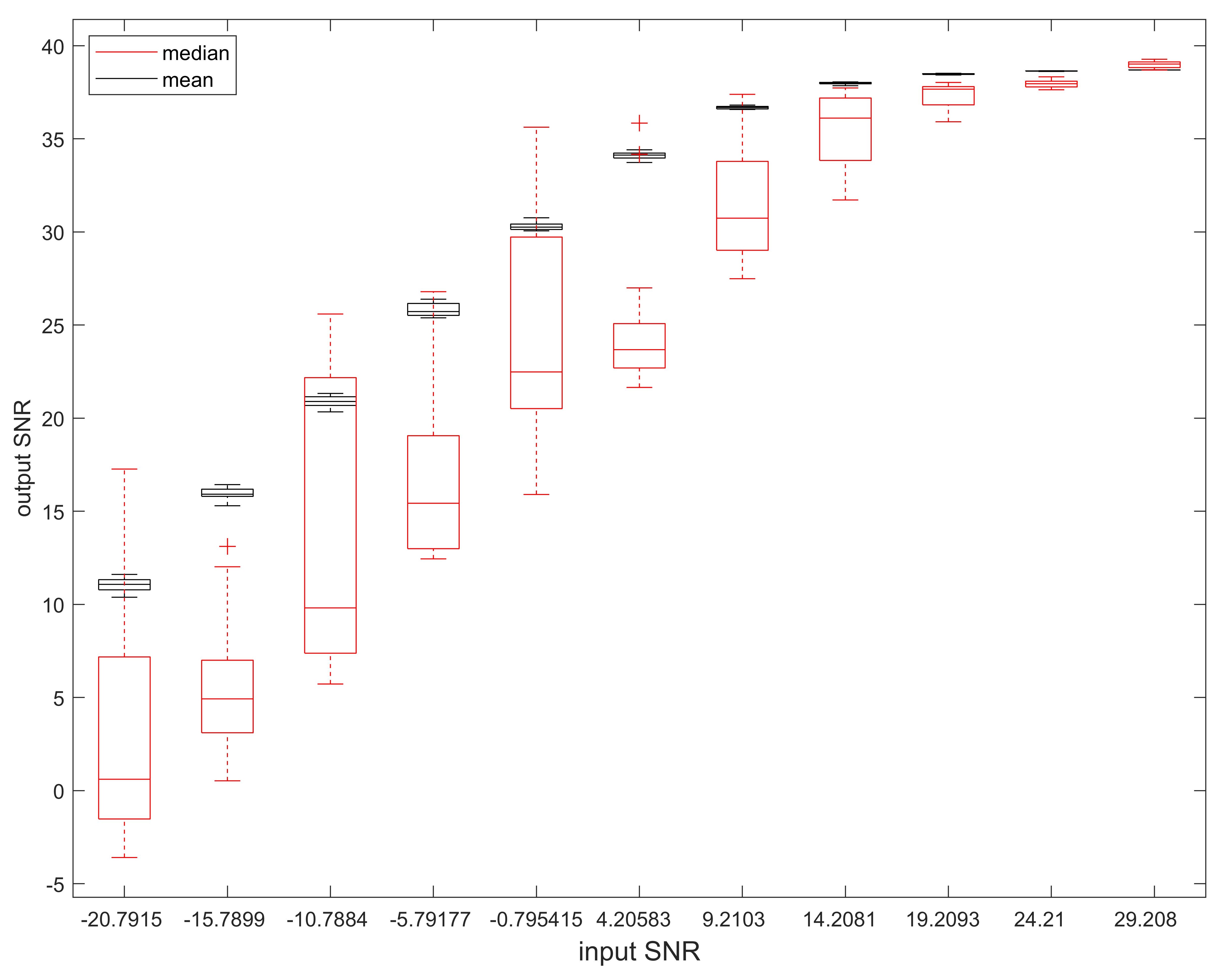}
}
\subfigure[]{
\includegraphics[width=3.3in]{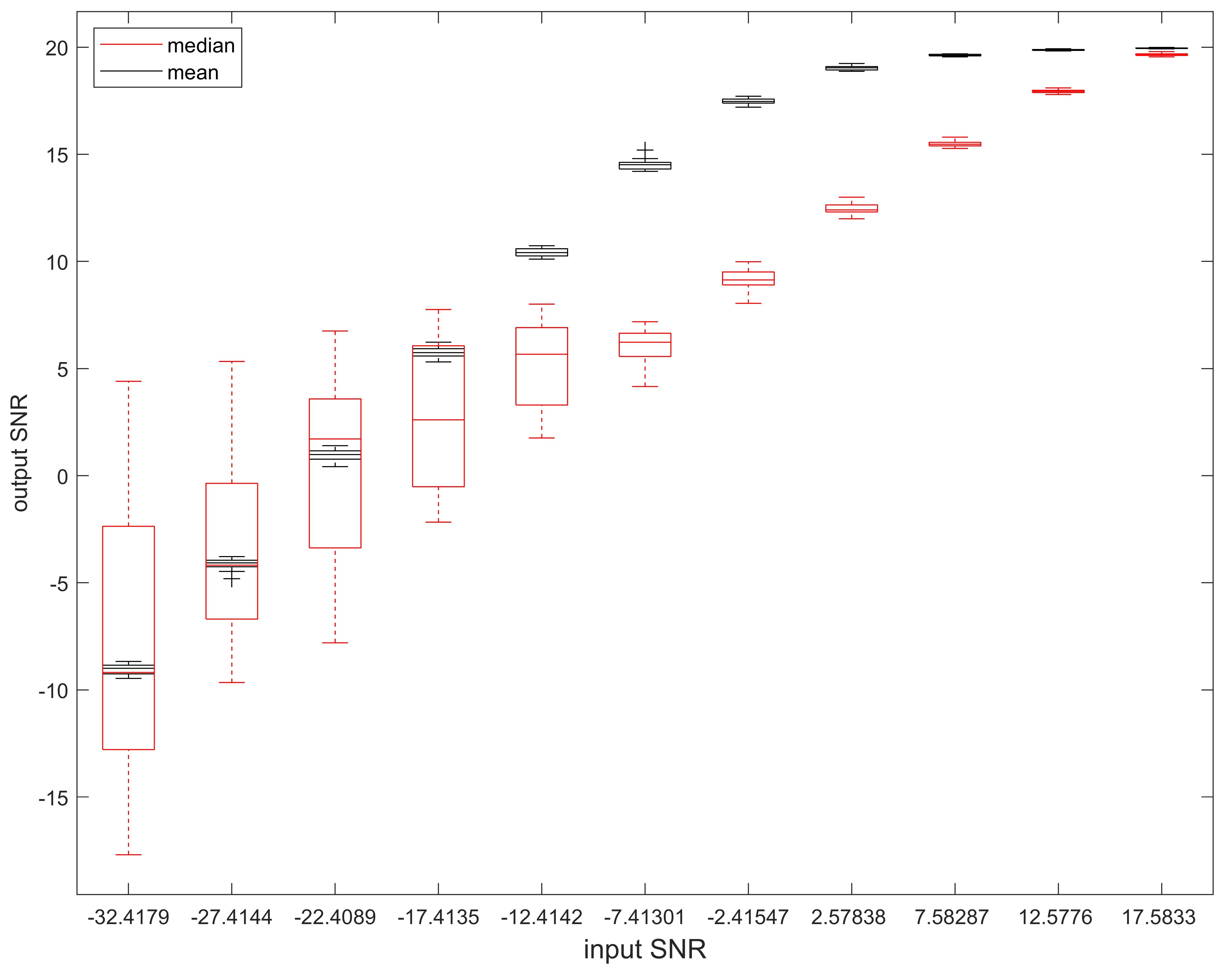}
}
\caption{The output SNR of graph mean and median filter versus input SNR. (a) The sea-level pressure sensor network dataset. (b) The sea-level temperature sensor network dataset. }

\end{figure}
\par Observe from the Fig. 4, we can obtain that the mean filter has a better performance, and the result of each filter is more stable than that of the median filter \cite{Ref21}.
\begin{figure}[H]
\centering
\subfigure[]{
\includegraphics[width=3.3in]{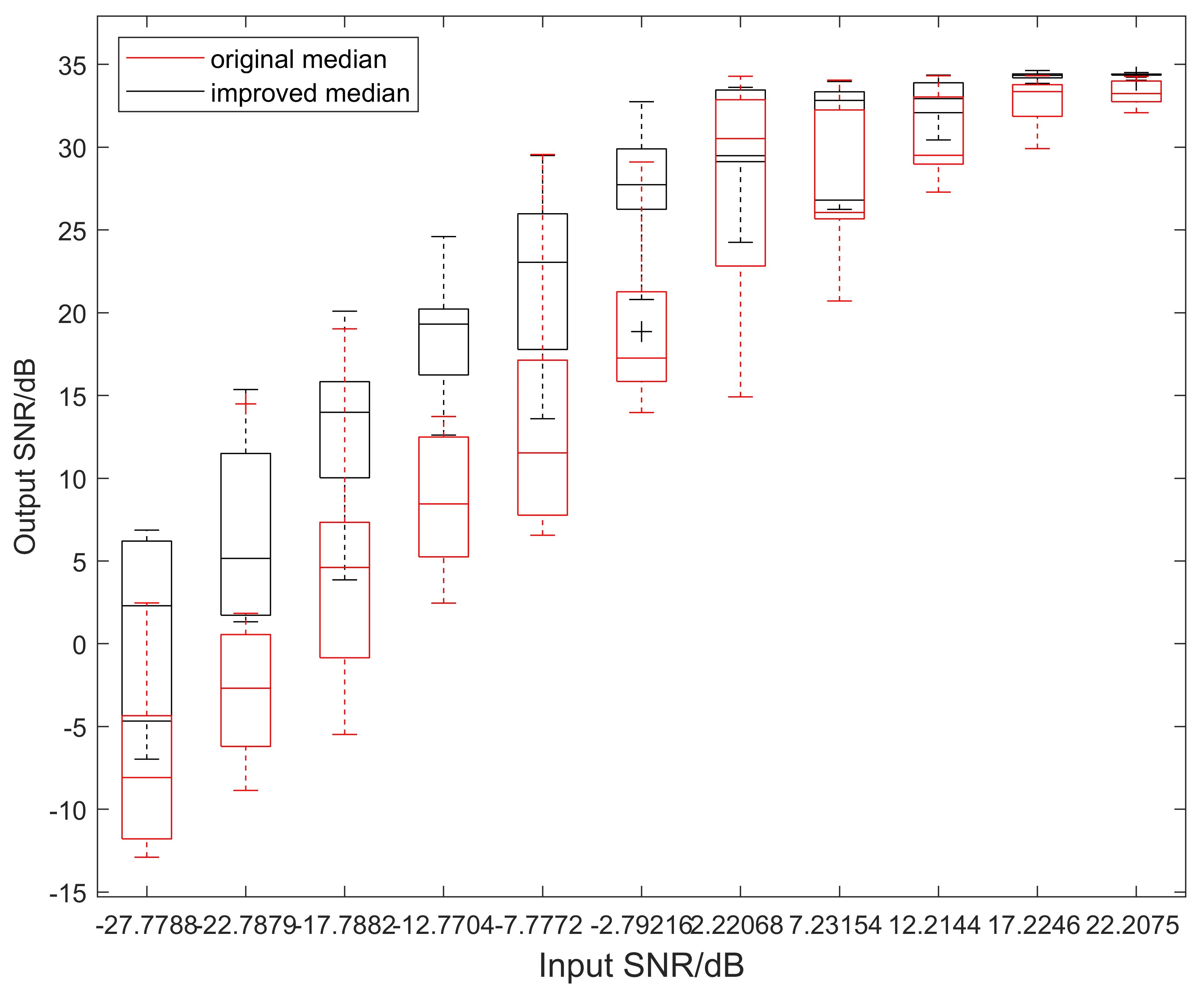}
}
\subfigure[]{
\includegraphics[width=3.3in]{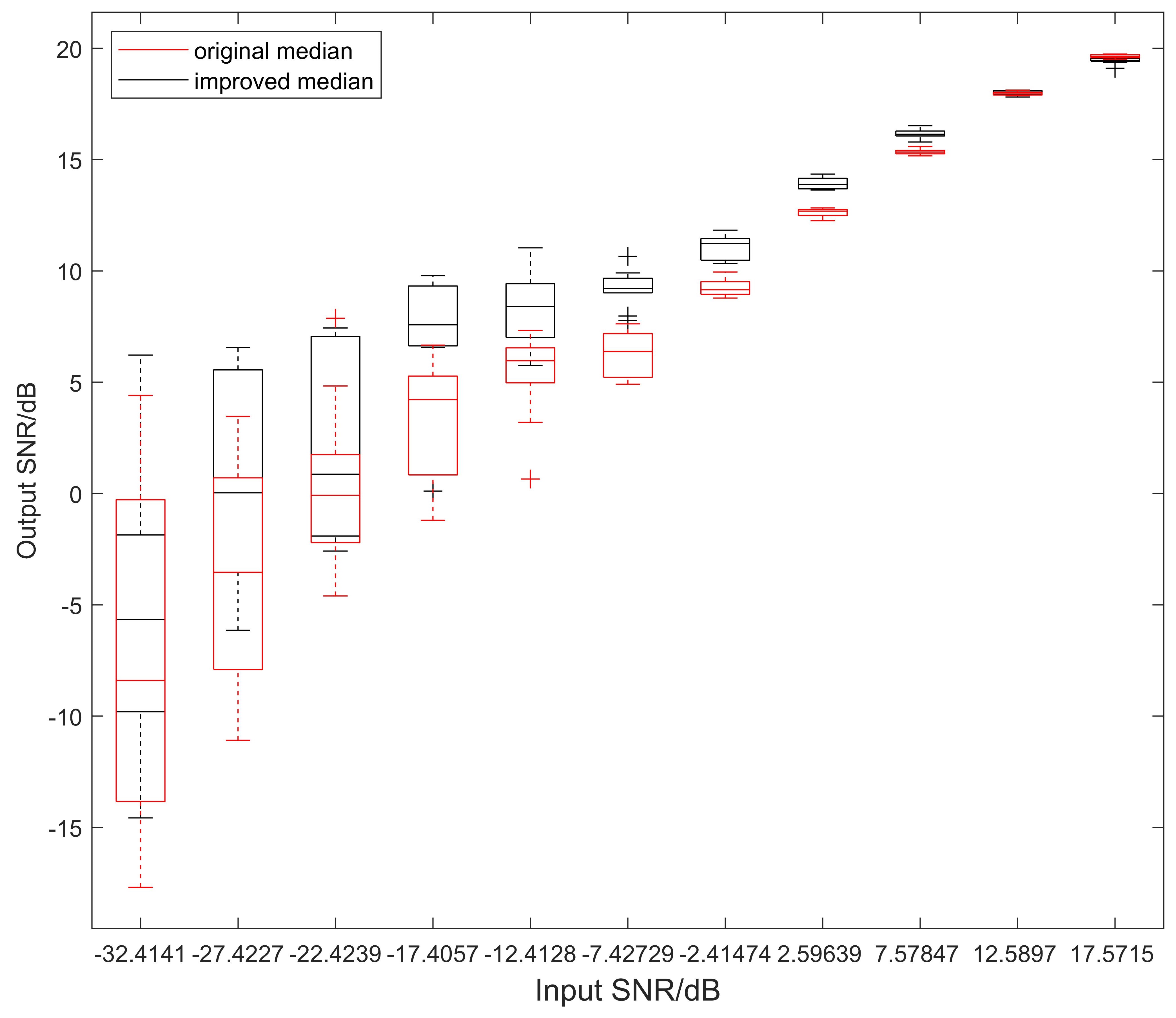}
}
\caption{The output SNR of graph median and improved median filter versus input SNR. (a) The sea-level pressure sensor network dataset. (b) The sea-level temperature sensor network dataset. }
\end{figure}
\par In Fig 5, (a) and (b) are the experimental results of sea level pressure sensor network and sea level temperature sensor network respectively. \emph{Original median} is the median filtering method proposed in \cite{Ref21}, and \emph{improved median} is the median filter method applying our node selecting graph. It can be seen from Fig.5 (a) that the application of our node selecting graph has better performance, especially at low SNR. Observe from Fig.2 (b), we can obtain that when the input signal noise is greater than -30dB and less than 12dB, the denoising performance of the median filter using our node selecting graph is obviously better than that of the original median filter.

\section{Conclusion}
In this paper, we propose a weighted attenuated K-Hop graph, and then we further propose a weighted node selection graph based on our K-Hop graph. According to the characteristics of white noise, we propose a mean graph filter based on this node selecting graph. Furthermore, we apply the node selectin graph to the median filter. The final experimental results show that our graph mean filters perform better in processing sensor network signals with white noise, and the experimental results are more stable. In addition, the experimental results show that the median filter using node selecting graph has better performance than the original median filter.

\bibliography{mybibfile}

\begin{thebibliography}{10}
\expandafter\ifx\csname url\endcsname\relax
  \def\url#1{\texttt{#1}}\fi
\expandafter\ifx\csname urlprefix\endcsname\relax\def\urlprefix{URL }\fi
\expandafter\ifx\csname href\endcsname\relax
  \def\href#1#2{#2} \def\path#1{#1}\fi

\bibitem{Ref1}
M.~Faria, C.~E. Cugnasca, J.~Amazonas, Insights into iot data and an innovative
  dwt-based technique to denoise sensor signals, IEEE Sensors Journal PP~(99)
  (2017) 1--1.

\bibitem{Ref2}
W.~Ye, S.~Li, X.~Zhao, A.~Abubakar, B.~Amine, A k times singular value
  decomposition based image denoising algorithm for dofp polarization image
  sensors with gaussian noise, IEEE Sensors Journal 18 (2018) 6138--6144.

\bibitem{Ref3}
X.~Chen, S.~Wu, C.~Shi, Y.~Huang, J.~Zhao, Sensing data supported traffic flow
  prediction via denoising schemes and ann: A comparison, Vol.~PP, 2020, pp.
  1--1.

\bibitem{Ref4}
D.~I. Shuman, S.~K. Narang, P.~Frossard, A.~Ortega, P.~Vandergheynst, The
  emerging field of signal processing on graphs: Extending high-dimensional
  data analysis to networks and other irregular domains, IEEE Signal Processing
  Magazine 30~(3) (2013) 83--98.

\bibitem{Ref5}
A.~Sandryhaila, J.~M. Moura, Big data analysis with signal processing on
  graphs: Representation and processing of massive data sets with irregular
  structure, Vol.~31, IEEE, 2014, pp. 80--90.

\bibitem{Ref6}
A.~Ortega, P.~Frossard, J.~Kova{\v{c}}evi{\'c}, J.~M. Moura, P.~Vandergheynst,
  Graph signal processing: Overview, challenges, and applications, Proceedings
  of the IEEE 106~(5) (2018) 808--828.

\bibitem{Ref7}
F.~Gama, A.~G. Marques, G.~Mateos, A.~Ribeiro, Rethinking sketching as
  sampling: A graph signal processing approach, arXiv preprint arXiv:1611.00119
  (2016).

\bibitem{Ref8}
E.~Isufi, P.~Banelli, P.~Di~Lorenzo, G.~Leus, Observing and tracking
  bandlimited graph processes from sampled measurements, Signal Processing 177
  (2020) 107749.

\bibitem{Ref9}
M.~Jorge Mendes~Spelta, W.~Alves~Martins, Normalized lms algorithm and
  data-selective strategies for adaptive graph signal estimation, Elsevier,
  2019.

\bibitem{Ref10}
N.~Perraudin, P.~Vandergheynst, Stationary signal processing on graphs, IEEE
  Transactions on Signal Processing 65~(13) (2017) 3462--3477.

\bibitem{Ref11}
A.~Sandryhaila, J.~M. Moura, Discrete signal processing on graphs, Vol.~61,
  IEEE, 2013, pp. 1644--1656.

\bibitem{Ref12}
A.~Sandryhaila, J.~M. Moura, Discrete signal processing on graphs: Frequency
  analysis, Vol.~62, IEEE, 2014, pp. 3042--3054.

\bibitem{Ref13}
E.~Isufi, A.~Loukas, A.~Simonetto, G.~Leus, Autoregressive moving average graph
  filtering, IEEE Transactions on Signal Processing 65~(2) (2016) 274--288.

\bibitem{Ref14}
X.~Shi, H.~Feng, M.~Zhai, T.~Yang, B.~Hu, Infinite impulse response graph
  filters in wireless sensor networks, IEEE Signal Processing Letters 22~(8)
  (2015) 1113--1117.

\bibitem{Ref15}
Z.~Xiao, H.~Fang, X.~Wang, Distributed nonlinear polynomial graph filter and
  its output graph spectrum: Filter analysis and design, IEEE Transactions on
  Signal Processing 69 (2021) 1--15.

\bibitem{Ref16}
Z.~Xiao, X.~Wang, Nonlinear polynomial graph filter for signal processing with
  irregular structures, IEEE Transactions on Signal Processing 66~(23) (2018)
  6241--6251.

\bibitem{Ref17}
S.~Segarra, A.~G. Marques, G.~R. Arce, A.~Ribeiro, Design of weighted median
  graph filters (2017) 1--5.

\bibitem{Ref18}
S.~Segarra, A.~G. Marques, G.~R. Arce, A.~Ribeiro, Center-weighted median graph
  filters (2016) 336--340.

\bibitem{Ref19}
F.~Grassi, A.~Loukas, N.~Perraudin, B.~Ricaud, A time-vertex signal processing
  framework: Scalable processing and meaningful representations for time-series
  on graphs, IEEE Transactions on Signal Processing 66~(3) (2017) 817--829.

\bibitem{Ref20}
D.~B. Tay, J.~Jiang, Time-varying graph signal denoising via median filters,
  IEEE Transactions on Circuits and Systems II: Express Briefs 68~(3) (2020)
  1053--1057.

\bibitem{Ref21}
D.~B. Tay, Sensor network data denoising via recursive graph median filters,
  Signal Processing 189 (2021) 108302.

\bibitem{Ref22}
T.~C. Aysal, K.~E. Barner, Generalized mean-median filtering for robust
  frequency-selective applications, IEEE Transactions on Signal Processing
  55~(3) (2007) 937--948.

\end{thebibliography}

\end{document}